\documentclass[prl,twocolumn,notitlepage,superscriptaddress,showkeys,showpacs]{revtex4-1}
\usepackage{amsmath,amsfonts,amssymb,amsthm,epsfig,array}
\usepackage{dsfont}
\usepackage{graphics}
\usepackage{float}
\usepackage{verbatim}
\usepackage[dvipsnames]{xcolor}
\usepackage[hidelinks,colorlinks,linkcolor=blue,
citecolor=blue,urlcolor=blue]{hyperref}
\usepackage{xr-hyper}
\usepackage[titletoc,title]{appendix}

\newcommand{\equaref}[1]{Eq.\;\eqref{#1}}
\newcommand{\figref}[1]{Fig.\;\ref{#1}}

\let\oldAA\AA
\renewcommand{\AA}{\text{\normalfont\oldAA}}

\newcommand\ket[1]{\left|#1\right>}

\makeatletter
\newcommand*{\addFileDependency}[1]{
  \typeout{(#1)}
  \@addtofilelist{#1}
  \IfFileExists{#1}{}{\typeout{No file #1.}}
}
\makeatother

\newcommand*{\myexternaldocument}[1]{%
    \externaldocument{#1}%
    \addFileDependency{#1.tex}%
    \addFileDependency{#1.aux}%
}
\myexternaldocument{supp}

\begin{document}
\title{Disordered Quantum Transport in Quantum Anomalous Hall Insulator-Superconductor Junctions }
\author{Jian-Xiao Zhang}
\affiliation{Department of Physics, the Pennsylvania State University, University Park, PA, 16802}
\author{Chao-Xing Liu}
\email{cxl56@psu.edu}
\affiliation{Department of Physics, the Pennsylvania State University, University Park, PA, 16802}
\begin{abstract}
In this communication, we numerically studied disordered quantum transport in a quantum anomalous Hall insulator-superconductor junction based on the effective edge model approach. In particular, we focus on the parameter regime with the free mean path due to elastic scattering much smaller than the sample size and discuss disordered transport behaviors in the presence of different numbers of chiral edge modes, as well as non-chiral metallic modes. Our numerical results demonstrate that the presence of multiple chiral edge modes or non-chiral metallic modes will lead to a strong Andreev conversion, giving rise to half-electron half-hole transmission through the junction structure, in sharp contrast to the suppression of Andreev conversion in the single chiral edge mode case. Our results suggest the importance of additional transport modes in the quantum anomalous Hall insulator-superconductor junction and will guide the future transport measurements.
\end{abstract}
\maketitle


{\it Introduction: }
The interplay between superconductivity proximity effect and chiral edge modes (CEMs) in a two-dimensional heterostructure with a quantum Hall (QH) or quantum anomalous Hall (QAH) system coupled to a superconductor (SC) has been a long-standing problem \cite{fu2008superconducting, fu2009josephson, qi2010chiral,takagaki1998conductance, takagaki1998transport, hoppe2000andreev, chtchelkatchev2007conductance,khaymovich2010andreev,sun2009quantum,akhmerov2007detection}. Recently, a strong resurgence of the research interest in this system results from the possible realization of topological superconductor (TSC) phase\cite{he2017chiral}. A TSC possesses a full bulk SC gap and gapless quasi-particle excitations, such as Majorana zero modes \cite{fu2008superconducting,sun2016majorana} or parafermions \cite{alicea2016topological,lindner2012fractionalizing,vaezi2014superconducting,mong2014universal,clarke2014exotic}, at the boundary. Non-Abelian statistics of these quasi-particle excitations enables the possibility of performing topological quantum computation based on TSC \cite{nayak2008non,alicea2011non}.

Early theoretical studies on the QH/SC junction focus on the Andreev reflection process occurring at the QH/SC interface \cite{takagaki1998conductance, takagaki1998transport, hoppe2000andreev, chtchelkatchev2007conductance,khaymovich2010andreev,sun2009quantum,akhmerov2007detection}
and the supercurrent flowing through CEMs\cite{van2011spin,zyuzin1994superconductor}.
Several early experiments on SC/semiconductor heterostructure have revealed evidence of Andreev reflections for two-dimensional electron gas in the high-Landau-level states under the magnetic fields.\cite{eroms2005andreev, moore1999andreev, batov2007andreev}.
Low-Landau-level states are challenging in these systems due to the limitation of the upper critical field of SCs and the high electron density\cite{wan2015induced}.
Recent experiments have shown that the low-Landau-level states at a relatively low magnetic field can be achieved in the graphene system by tuning electron density through gate voltages, thus leading to significant progress in inducing SC correlation via the proximity effect into the graphene in the QH regime. Transport evidences, including supercurrents carried by CEMs\cite{amet2016supercurrent,calado2015ballistic}, enhanced QH plateau conductance due to Andreev process \cite{rickhaus2012quantum}, cross Andreev conversion (AC)\cite{cayssol2008crossed} and inter-Landau-level Andreev reflection \cite{sahu2018inter}, have been found in graphene QH systems in contact to SC electrodes under an external magnetic field. These encouraging experimental progresses lay the foundations for the theoretical proposals of realizing chiral Majorana modes (CMMs)\cite{qi2010chiral, chung2011topological} and Majorana zero modes\cite{chen2018quasi,alicea2012new,zeng2018quantum}
in the SC/QH (or QAH) junctions.

More recently, an observation of $e^2/2h$ conductance kink is claimed to be the transport evidence of CMMs in the TSC phase of a QAH/SC hetero-structure\cite{he2017chiral}.
However, this claim is still under debates \cite{kayyalha2019non, zhang2019note} because the early theoretical prediction was based on the calculation of the clean QAH/SC hetero-structure model with the Landauer-Buttiker formalism \cite{chung2011topological}
while the QAH samples in experiments, particularly the SC/QAH interface, are highly disordered \cite{ji20181,huang2018disorder,lian2018quantum}. The mean free path $l_{\textrm{mfp}}$ around the SC/QH(QAH) interface region is greatly reduced due to interface roughness and normally much smaller than the typical length scale of the SC/QH(QAH) interface. For example, in Ref. \cite{lee2017inducing}, $l_{\textrm{mfp}}$ is around 0.3nm, much smaller than the width of SC ribbon (around $50\sim 600$nm). Theoretically, the disorder effect in the SC/QH interface was investigated for the high-Landau-level systems based on either the semi-classical skipping orbit picture or the Landau level picture in the early literature \cite{takagaki1998conductance, takagaki1998transport,chtchelkatchev2007conductance,khaymovich2010andreev}.
More recently, several theoretical models are developed for the disorder-induced bulk topological phase transition in the QAH/SC hetero-structures \cite{huang2018disorder,ji20181,lian2018quantum,kramer2005random,chalker2001thermal}. The disorder effect from elastic scattering is normally not important for CEM transport in the QH or QAH regime. However, the conductance oscillation can be induced by AC of CEMs propagating through the SC/QH (or QAH) interface in the ballistic regime \cite{lian2016edge,gamayun2017two}, which is sensitive to elastic scattering. Therefore, understanding disorder effect in the SC/QH(QAH) junctions \cite{wang2018multiple,lian2019distribution} is essential for any reliable theoretical interpretation.

\begin{figure}[htp]
\centering
\includegraphics[width=\columnwidth]{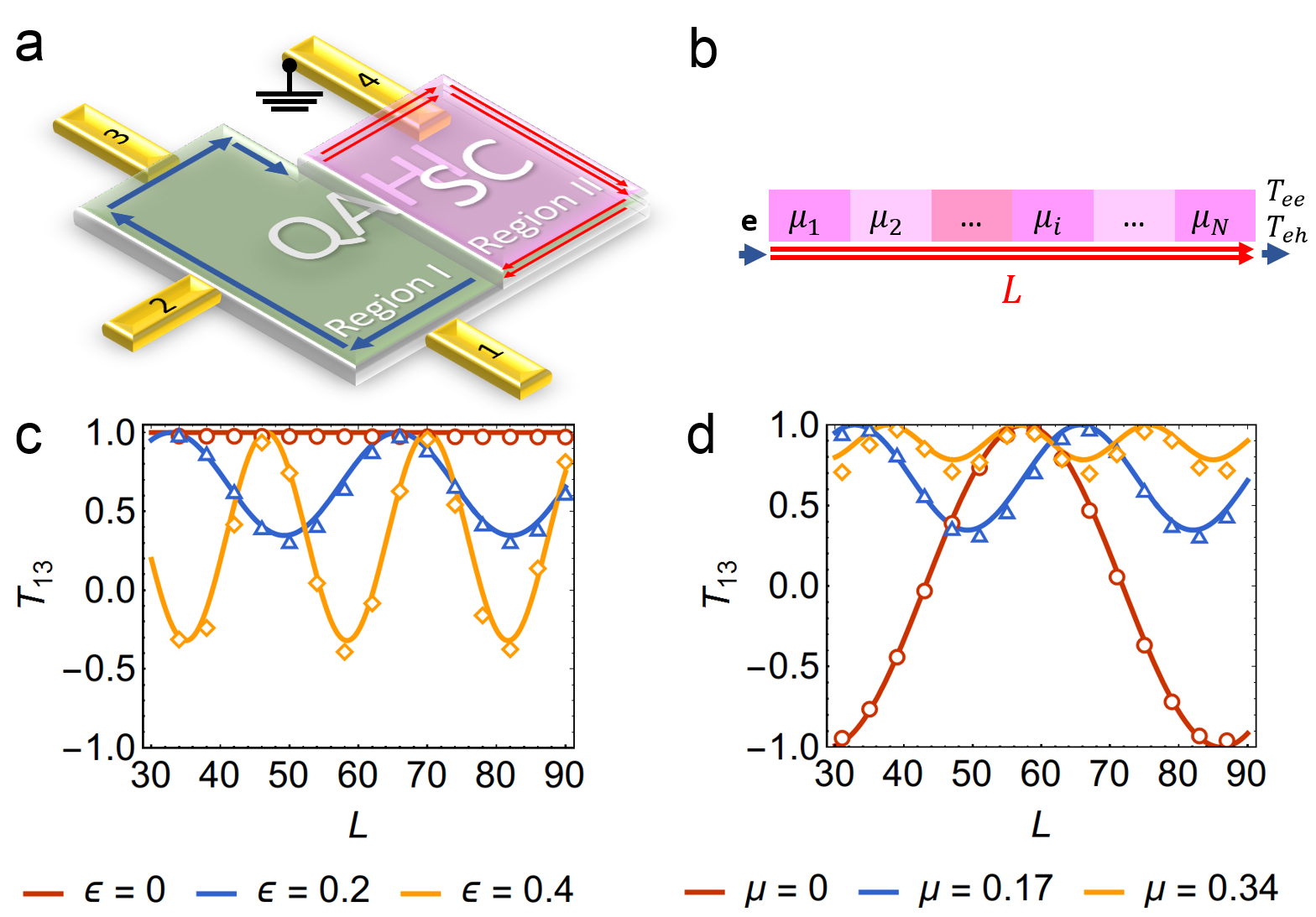}
\caption{(a) A schematics of the configuration for the SC-QAHI heterostructure with four leads labelled by 1 to 4.
(b) The trajectory of chiral edge modes going through the region II in (a), which is segmented into multiple regions with local chemical potential $\mu_i$ ($i=1,...,N$).
(c) The oscillation of $T_{13}$ at a fixed chemical potential $\mu_i=\mu = 0.1$ for different $\epsilon$ in the clean limit.
(d) The oscillation of $T_{13}$ at finite electron energy $\epsilon = 0.2$ in the clean limit $\mu_i = \mu$ for all $i$. The circles, triangles and diamonds are calculated from the microscopic model while the solid lines are from effective edge model.
}
\label{fig:scheme}
\end{figure}

In this work, we focus on the disordered transport through the QAH/SC junction. We consider a theoretical model for the setup with SC partially covering the QAH system, forming a planar junction between a pure QAH region (Region I) and a SC/QAH vertical junction region (Region II), as shown in \figref{fig:scheme}a.
We assume SC proximity effect is weak in the QAH system and thus the region II is topologically equivalent to the QAH phase. As a result, the CEM at the boundary of the region I is transformed into two CMMs along the boundary of the region II (\figref{fig:scheme}a). Previous theoretical studies on the similar configurations have revealed conductance oscillation\cite{lian2016edge,gamayun2017two} in the ballistic transport regime. We focus on the elastic-scattering-dominated transport regime, in which the edge length $L$ of the QAH/SC junction is much larger than the mean free path $l_{\textrm{mfp}}$.
In this transport regime, the disordered transport behavior sensitively depends on the number $N$ of CEMs and the existence of other transport channels. In the single CEM ($N=1$) case, we find that the transmission through the boundary of QAH/SC junction region approaches unity without AC even in the presence of disorders when the incident electron energy approaches 0 (or the zero-bias limit in experiments). This is a consequence of the $p$-wave nature of the allowed SC proximity effect in the single CEM case. In contrast, the transmission will quickly decay away from one and reach certain saturating values when considering the finite incident electron energy (or the finite bias), or multiple CEMs ($N>1$), or the coexistence of a CEM and a non-chiral metallic mode. These results bring new insights into the existing experiments of QAH/SC junction and reveal the important role of the interplay between elastic scattering and CMM transport.

{\it Model Hamiltonian and Transport of a single CEM: }
Due to topological equivalence between the QH and QAH states, we consider a two-band model of QAH insulator and couple it to a SC. The Bogoliubov-de Gennes (BdG) Hamiltonian for a QAH/SC junction can be written as
\begin{equation}
H=\begin{pmatrix} H_{\textrm{QAH}} & H_{\Delta} \\ H_{\Delta}^\dagger & -H_{\textrm{QAH}}^* \end{pmatrix}
\label{eq:Hamiltonian}
\end{equation}
with $H_{\textrm{QAH}}=(M + B (\partial_x^2+\partial_y^2))\sigma_z+\alpha\sigma_x(-i\partial_x)+ \alpha\sigma_y (-i\partial_y) - \mu$ and $\sigma_{x,y,z}$ are the Pauli matrices for spin \cite{lian2016edge}. We choose the superconducting gap $H_\Delta = i \sigma_y (\Delta_0 + \Delta_z (-i \partial_x))$ with both spin-singlet component $\Delta_0$ and triplet component $\Delta_z$. The reason to include triplet components is because only triplet SC is allowed for the single CEM case, as discussed in the edge theory below. The parameters $\mu$, $\Delta_0$ and $\Delta_z$ are in general spatially dependent, to incorporate the spatial configuration of junctions and disorder-induced variations.
We apply the Hamiltonian (\ref{eq:Hamiltonian}) to the configuration in \figref{fig:scheme}a on a retanglar geometry
and adopt the Recursive Green's function method combined with Landauer-B\"uttiker formalism (see Appendix A and Appendix F for details). The leads 1, 2 and 3 are attached on the edge of the sample, while the lead 4 is connected to the SC bulk and grounded.

We first study the transport behavior of the Hamiltonian (\ref{eq:Hamiltonian}) in the clean limit and consider the transmission from the lead 3 to 1, labeled as $T_{13}$, which accounts for the transmission through two CMMs at the boundary of the region II (see \figref{fig:scheme}a).
Through this region, electrons can transmit as an electron or convert to a hole through the AC process. We denote the electron-electron transmission as $T_{ee}$ and the electron-hole transmission of AC as $T_{eh}$, thus $T_{13} = T_{ee} - T_{eh}$. In \figref{fig:scheme}c and d, the calculated $T_{13}$ as a function of the length $L$ is shown as circles, triangles and diamonds for different incident electron energies $\epsilon$ and different chemical potentials $\mu$, respectively. The oscillation behavior of transmission comes from the AC at the SC/QAH boundary \cite{lian2016edge,gamayun2017two}. The amplitude of the oscillation increases with $\epsilon$ (or Fermi momentum of CEMs), but decreases with $\mu$. In the $\epsilon=0$ case, the transmission $T_{13}$ always keeps 1, suggesting the suppression of AC in the zero-bias limit, which is consistent with the literature\cite{van2011spin}.

Since we focus on the edge transport regime with the chemical potential within the bulk gap of $H_{\textrm{QAH}}$, the full Hamiltonian of the QAH/SC junction can be projected into the subspace spanned by CEMs, giving rise to
\begin{equation}
H_\textrm{eff}=
\begin{pmatrix} v_f k - \mu		& -v_\Delta k \\
							-v_\Delta k 			    & v_f k + \mu
\end{pmatrix}
\label{eq:H_eff}
\end{equation}
on the CEM basis of $\ket{\phi_{e}}$ and $\ket{\phi_{h}}$, where $v_f$ is the Fermi velocity of CEM and $v_\Delta$ is the coefficient of triplet pairing component. Detailed derivation of the effective model can be found in Appendix B. We notice that only the triplet component can contribute to the pairing term for CEMs \cite{van2011spin}, as guaranteed by the particle-hole symmetry. As described in Appendix C, the transmission $T_{13}$ can be directly computed through the scattering matrix approach and given by
\begin{equation}
T_{13}=
1-\frac{2\epsilon^2 v_\Delta^2 \sin^2\left(\frac{\sqrt{\epsilon^2 v_\Delta^2 + (v_f^2-v_\Delta^2)\mu^2}}{v_f^2-v_\Delta^2}L\right)}{\epsilon^2 v_\Delta^2 + (v_f^2-v_\Delta^2)\mu^2},
\label{eq:p_Transmission}
\end{equation}
from which one finds $T_{13}$ oscillates with the amplitude $\frac{2\epsilon^2 v_\Delta^2 }{\epsilon^2 v_\Delta^2 + (v_f^2-v_\Delta^2)\mu^2}$ and the period $\frac{v_f^2-v_\Delta^2}{2\sqrt{\epsilon^2 v_\Delta^2 + (v_f^2-v_\Delta^2)\mu^2}}$. The oscillation amplitude increases when increasing $\epsilon$, but decreases when increasing $\mu$, while the oscillation period decreases with increasing either $\epsilon$ or $\mu$. All these features are consistent with the numerical simulations above. In addition, as the incident electron energy $\epsilon$ approaches zero, the oscillation disappears according to \equaref{eq:p_Transmission}, and thus the transmission $T_{13}$ always stays at 1, independent of chemical potential $\mu$. Physically, this behavior originates from the $p$-wave nature of the pairing in the effective edge model.  By choosing appropriate parameters in Eq. \ref{eq:p_Transmission}, the calculated transmission $T_{13}$ (solid lines in \figref{fig:scheme}c,d) can fit well with that from the numerical simulations of the full model (circles, triangles and diamonds in \figref{fig:scheme}c,d),
thus justifying the validity of the effective edge Hamiltonian (\ref{eq:H_eff}).

We next consider the influence of disorder scattering on the transmission $T_{13}$, particularly the on-site fluctuation of chemical potential $\mu$, based on the effective edge Hamiltonian (\ref{eq:H_eff}). To do that, we divide the QAH/SC interface into $N$ segments with the chemical potential $\mu_i$ ($i=1,2...,N$) chosen to be a random variable from a uniform distribution on $[-\mu_{\textrm{imp}}/2,\mu_{\textrm{imp}}/2]$ (See \figref{fig:scheme}b). For each disorder configuration $\{ \mu_i\}_{i=1,...,N}$, we compute the transmission $T_{13}$ through the transfer matrix formalism. The physical transmission, denoted as $\bar{T}_{13}$, is obtained by averaging multiple independent disorder configurations.

\begin{figure}[htp]
\centering
\includegraphics[width=\columnwidth]{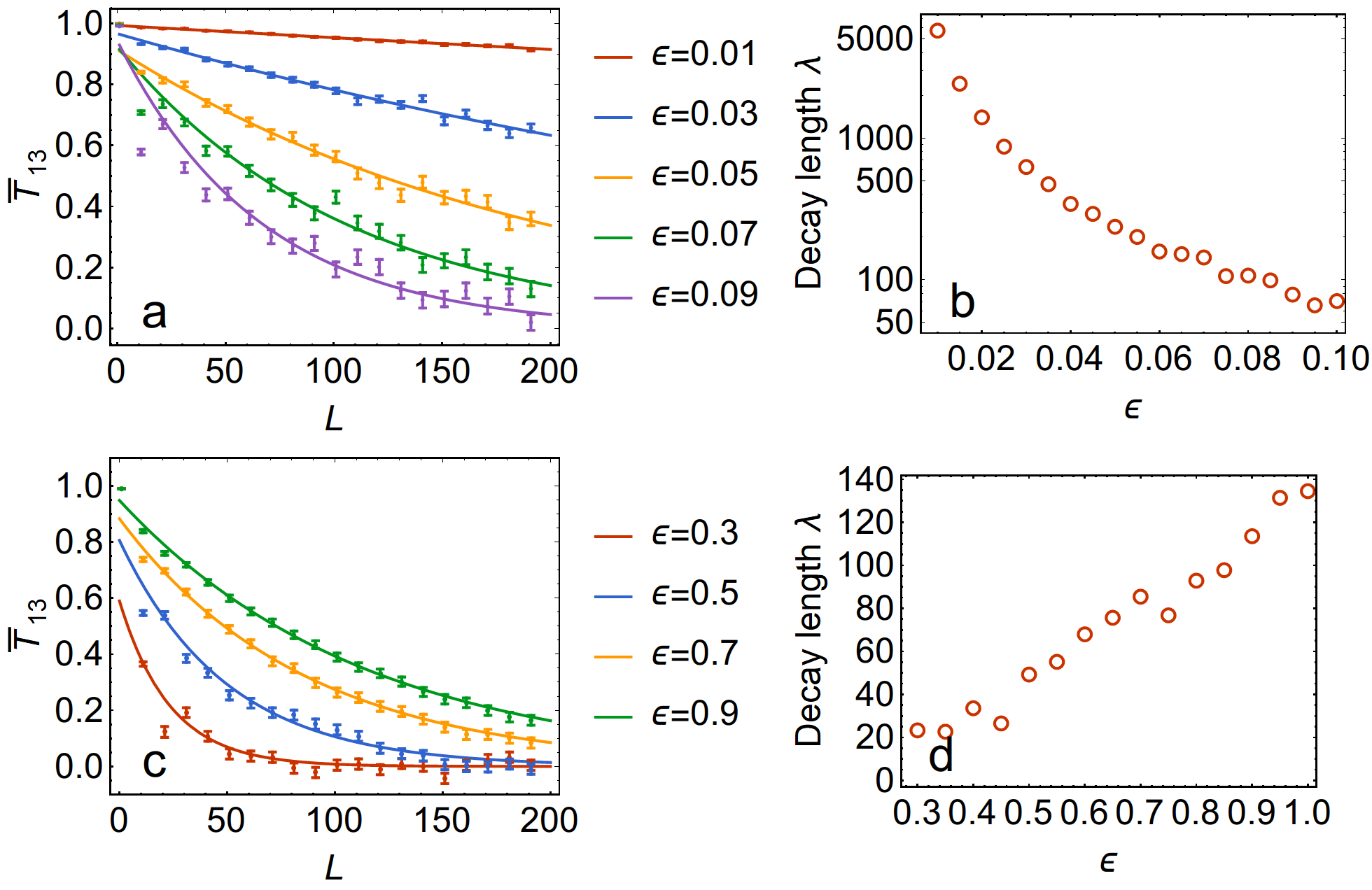}
\caption{(a, b) \textit{Single MEM case:} (a) Average transmission $\bar{T}_{13}$ as a function of L for different $\epsilon$ with $\mu=0.1$. The error bar depicts the standard error over 1000 configurations. (b) Decay length $\lambda$ extracted from transmission as a function of $\epsilon$. Other parameters are $v_f=1, v_\Delta=0.5, \mu_\textrm{imp} = 0.5$. (c, d) same as (a, b) but for the \textit{Multiple MEMs case}. . Other parameters for are  $v_{f1}=1, v_{f2} = 2, \Delta=0.1, \mu_\textrm{imp} = 1$. The fundamental difference between the two cases is the divergence of decay length at zero energy in the single MEM case.}
\label{fig:psModel}
\end{figure}

\figref{fig:psModel}a shows that the averaged transmission $\bar{T}_{13}$ decays exponentially with respect to the length $L$. The decay behavior of $\bar{T}_{13}$ can be understood as the decoherent interference between different trajectories of electron-hole oscillation as varying chemical potential.
The decay length, denoted as $\lambda$, can be extracted from \figref{fig:psModel}a and its dependence on $\epsilon$ is shown in \figref{fig:psModel}b. As $\epsilon$ approaches zero, $\lambda$ increases rapidly, and thus $\bar{T}_{13}$ almost remains 1 when increasing $L$, indicating the suppression of AC in this limit, even when including disorder scattering. The dependence of $\bar{T}_{13}$ on $\epsilon$ and $\mu$ for a fixed $L$ is discussed in Appendix D.

{\it Disordered transport of multiple modes at the QAH/SC interface: }
In real experiment devices, multiple transport channels may exist at the QAH/SC interface, including (1) high-Landau level states with multiple CEMs ($N>1$) for the QH states \cite{amet2016supercurrent,cayssol2008crossed} and (2) the coexistence of a CEM and other non-chiral metallic modes \cite{wang2013anomalous,chang2015zero}. Therefore, we next go beyond the single CEM case and study the influence of multiple modes on disordered transport.

\begin{figure}[htp]
\centering
\includegraphics[width=0.8\columnwidth]{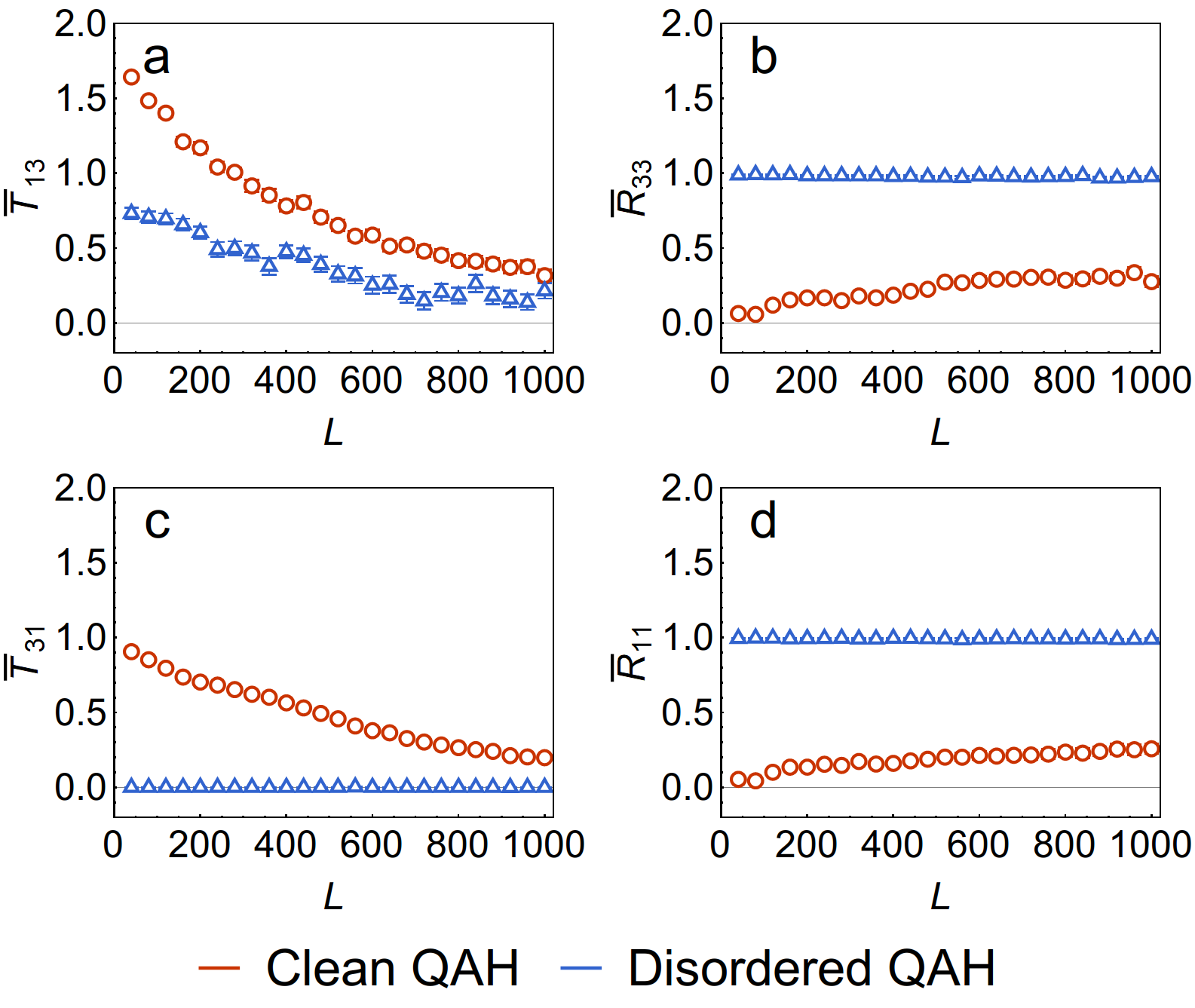}
\caption{Transmissions (a, c) and reflections (b, d) between lead 1 and 3 as functions of system length, for the case with both CEM and non-chiral modes. Two different scenarios where the QAH region (region I) is clean (red) and disordered (blue) are marked.
Chemical potential values are $\mu_1 = -0.3, \mu_2 = -3.2$.
For other parameters, please see Appendix E.
}
\label{fig:6band}
\end{figure}

We first consider the case with two CEMs ($N=2$) for simplicity. We denote the basis of the effective Hamiltonian as $\{ \ket{e_1}, \ket{e_2}, \ket{h_1}, \ket{h_2} \}$, where the subscript labels two channels. In addition to the $p$-wave pairing between $\ket{e_{1(2)}}$ and $\ket{h_{1(2)}}$, the pairing between $\ket{e_{1(2)}}$ and $\ket{h_{2(1)}}$ also exists and can be of $s$-wave nature. Therefore, if we only keep the lowest order terms, the effective Hamiltonian can be written as
\begin{equation}
\begin{split}
H_\textrm{m}&=
    \begin{pmatrix}
     v_{f1} k-\mu _1 & 0 & 0 & \Delta  \\
     0 & v_{f2} k-\mu _2 & -\Delta  & 0 \\
     0 & -\Delta  & v_{f1} k+\mu _1 & 0 \\
     \Delta  & 0 & 0 &  v_{f2} k + \mu _2 \\
    \end{pmatrix}
\end{split}
\label{eq:H_mult}
\end{equation}
with the pairing term $\Delta$.
Transport simulations can be performed for the Hamiltonian (\ref{eq:H_mult}) with the on-site chemical potential disorders. \figref{fig:psModel}c shows the averaged transmission $\bar{T}_{13}$ as a function of $L$, which shows an exponential decay with the decay length $\lambda$ depending on $\epsilon$ in \figref{fig:psModel}d. In sharp contrast to the single-CEM case, the decay length increases with $\epsilon$ and is generally quite small for the multiple-CEMs case. As a consequence, the transmission $\bar{T}_{13}$ is always close to zero for a range of $\epsilon$ and $\mu$ when the length $L$ becomes large. (See Fig.\;D.1c,d in Appendix D.
Physically, the constant $\Delta$ term can induce a large oscillation in the clean limit and thus disorder can induce a strong dephasing of the oscillation pattern, giving rise to the decay behavior. This conclusion is consistent with the previous studies on the transport through the QH/SC interface for the high-Landau-level QH state \cite{takagaki1998conductance, takagaki1998transport,chtchelkatchev2007conductance,khaymovich2010andreev}.

We next consider the coexistence of the CEM and the non-chiral metallic mode, and this scenario has been theoretically proposed \cite{wang2013anomalous} and later experimentally demonstrated \cite{chang2015zero} in magnetically doped TI films. We consider the BdG Hamiltonian
\begin{equation}
H_\textrm{orig}=
\begin{pmatrix}
 H_{\textrm{elec}}(k) & H_\Delta  \\
 H_\Delta^\dagger  &  -H_{\textrm{elec}}^*(-k), \\
\end{pmatrix}\label{eq:H_hybr}
\end{equation}
where
\begin{equation}
H_{\textrm{elec}} = \begin{pmatrix}
v_{f} k-\mu_1 & \xi_1 & \xi_2 \\
\xi_1  &  v_{1} k - \mu_2  & 0\\
\xi_2  &  0  & -v_{1} k - \mu_2
\end{pmatrix}
\label{eq:H_elec}
\end{equation} and
\begin{equation}
H_\Delta = \begin{pmatrix}
0 & \Delta_1 & \Delta_2 \\
-\Delta_1  &  0  &  0\\
-\Delta_2  &  0  &  0
\end{pmatrix}.
\end{equation}

Here $H_{\textrm{elec}}$ is written on the basis $\{ \ket{e_c}, \ket{e_{nL}}, \ket{e_{nR}} \}$, where $\ket{e_c}$ labels the CEM while $\ket{e_{nL}}$ and $\ket{e_{nR}}$ together represent the non-chiral mode due to their opposite velocities. In magnetic TI films, non-chiral modes originate from the quasi-helical gapless modes at the side surfaces of TI films \cite{wang2013anomalous}. The coupling between the CEM and non-chiral mode is described by the parameters $\xi_{1,2}=\xi$. The $s$-wave pairing gap can exist between the CEM and non-chiral mode, chosen as $\Delta_{1,2}=\Delta$ in $H_\Delta$, making this system similar to the multiple CEMs case. On the other hand, the non-chiral nature also suggests the existence of backward propagating channel and thus will strongly affect the transport behaviors in the QAH system \cite{wang2013anomalous}.
To understand its influence, we below discuss two different scenarios, both may occur in actual experiments. The details of the edge models can be found in Appendix E, with the transmissions/reflections of both models shown in Fig.\;E.3 and Fig.\;E.4.

In the first scenario, we assume disorders exist in the QAH/SC junction (region II) while the transport in the QAH side (region I) is ballistic (or quasi-ballistic), later referred as the ``clean QAH'' case. In this situation, the non-chiral modes have been experimentally shown to induce non-local transport signal in the QAH system \cite{chang2015zero}. In our setup, the ballistic transport of non-chiral modes in region I leads to a negligible backscattering for $\bar{R}_{33}$ and $\bar{R}_{11}$ at small $L$, both saturating to certain values when increasing $L$ in \figref{fig:6band}b. The transmissions $\bar{T}_{13}$ and $\bar{T}_{31}$ starts from 2 and 1 for a small $L$, respectively, reflecting the number of the left and right moving modes. For a large $L$, both transmissions decay to zero in \figref{fig:6band}a. The existence of non-zero $\bar{R}_{33}$ and $\bar{R}_{11}$ makes this situation quite different from other situations.

For the second scenario, disordered transport is assumed for the whole QAH insulator, spanning over both regions I and II in \figref{fig:scheme}a, and later referred as the ``disordered QAH'' case. Due to the Anderson localization, the 1D non-chiral mode is completely localized, as manifested by the reflection $\bar{R}_{33}\approx \bar{R}_{11}\approx 1$ and $\bar{T}_{31}\approx 0$ in \figref{fig:6band}b, c and d. Even though getting localized, the non-chiral mode still mediates the AC at the QAH/SC interface (region II in \figref{fig:scheme}a). Consequently, the transmission $\bar{T}_{13}$ exponentially decays to zero for a large $L$, even when $\epsilon=0$, making this situation more similar to the multiple CEMs case.

\begin{figure}[htp]
\centering
\includegraphics[width=\columnwidth]{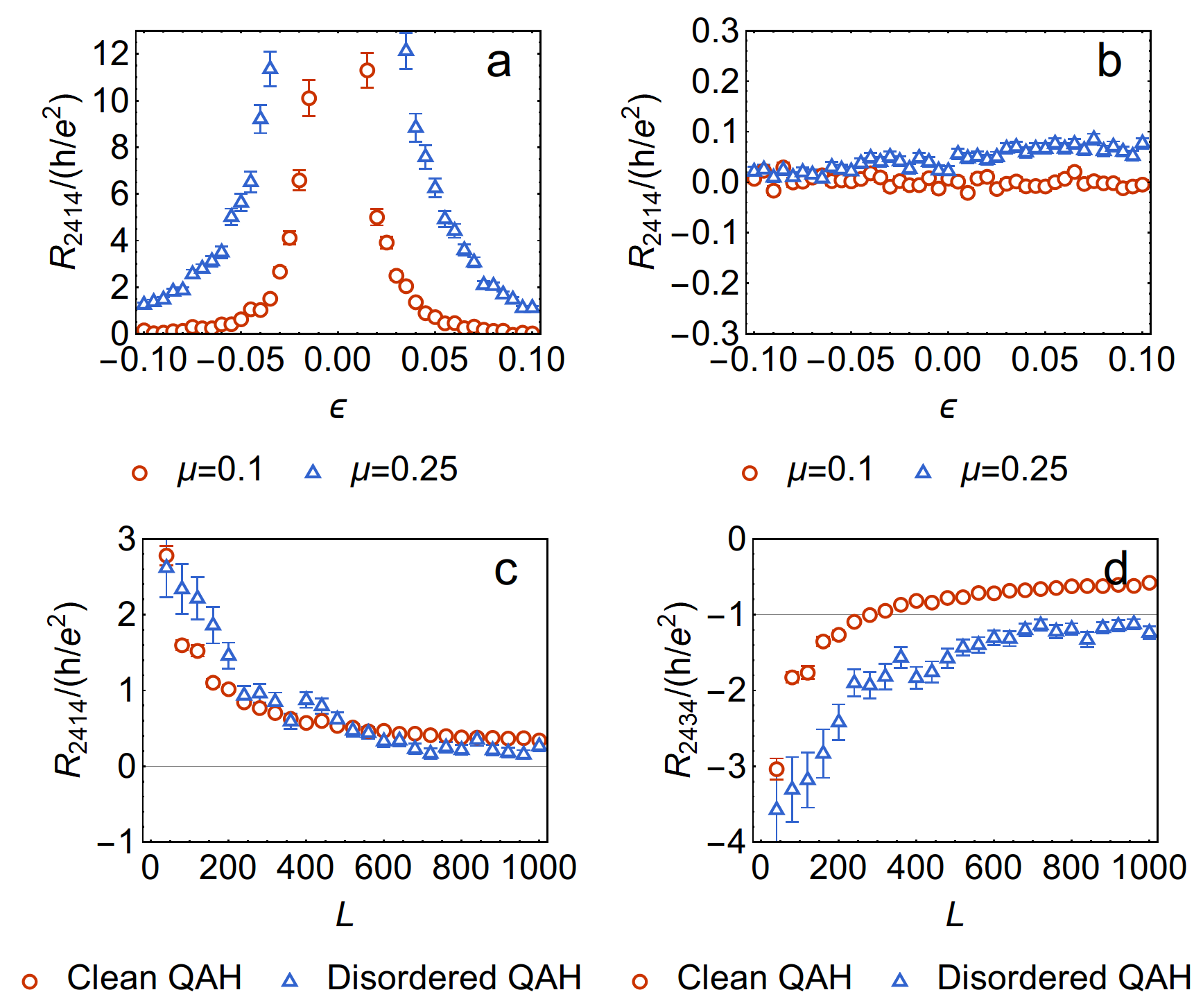}
\caption{(a,b) Resistances $R_{24,14}$ as a function of $\epsilon$ for the (a) single CEM case and (b) multiple CEMs case. System length $L = 200$. Red circles and blue triangles stand for different chemical potential values. (c, d) Resistances $R_{24,14}$ and $R_{24,34}$ as a function of $L$ for the coexistence of CEM and metallic mode system, with red circles and blue triangles representing the clean QAH and the disordered QAH cases, respectively. For disordered QAH, the values of $R_{24,14}$ and $R_{24,34}$ saturate to 0 and $-h/e^2$, respectively. The parameters used for c and d are $\mu_1 = -0.3, \mu_2 = -3.2$.}
\label{fig:resistance}
\end{figure}

{\it Experimental Relevance: }
Finally, we examine the behaviors of resistance in the experimental setup of \figref{fig:scheme}a. 
We consider the current driven from the leads 2 to 4 and discuss the resistance $R_{24,14} \equiv V_{14}/I_{24}$ and $R_{24,34}\equiv -V_{34}/I_{24}$ for four different cases: (i) the single CEM case, (ii) the multiple CEMs case, (iii) the coexistence of the CEM and non-chiral metallic mode, with a clean QAH region and (iv) with a disordered QAH region. The case (i) shows a unique insulating behavior for $\epsilon\sim 0$ in \figref{fig:resistance}a, while $R_{24,14}$ and $R_{24,34}$ are insensitive to the energy $\epsilon$ for all the other cases (see \figref{fig:resistance}b and Fig.\;E.5a and b in Appendix E).
For the cases (i) and (iv), $R_{24,34}$ and $R_{24,14}$ are related by $R_{24,14} + R_{24,34} = -h/e^2$, while for the case (ii), $R_{24,14}$ by $R_{24,14} + R_{24,34} = -h/2e^2$ (see \figref{fig:resistance}c and d, and Fig.\;E.5 in Appendix E). For the case (iii), due to ballistic transport of non-chiral metallic mode, $R_{24,14}$ and $R_{24,34}$ are independent of each other in \figref{fig:resistance}c and d. Therefore, measuring $R_{24,14}$ and $R_{24,34}$ simultaneously can distinguish these different cases.

{\it Acknowledgement: }
We are thankful for the helpful discussion with Biao Lian and Jiabin Yu.
We acknowledge the support of the Office of Naval Research (Grant No. N00014-18-1-2793), the U.S. Department of Energy (Grant No.~DESC0019064) and Kaufman New Initiative research grant KA2018-98553 of the Pittsburgh Foundation.

\bibliography{DisorderChiralMajorana}

\end{document}


\title{Supplementary Materials for ``Quantum transport of Chiral Majorana Modes in Disordered system''}

\author{Jian-Xiao Zhang}
\affiliation{%
 Department of Physics, The Pennsylvania State University, University Park, Pennsylvania 16802-6300, USA
}%
\author{Chao-Xing Liu}%
\affiliation{%
 Department of Physics, The Pennsylvania State University, University Park, Pennsylvania 16802-6300, USA
}

\maketitle
\appendix
\renewcommand\thefigure{\thesection.\arabic{figure}}
\renewcommand{\theequation}{\thesection.\arabic{equation}}

\section{Numerical Green's function method for transport calculation\label{app:numerical}}

The transmission is calculated from the standard Green's function method combined with the Landauer-B\"uttiker formalism.  \cite{datta1997electronic} As the system possesses the particle-hole symmetry, the conservation of charge in a realistic system requires the bulk of the superconductor to be grounded, shown as lead 4 in Fig.\;1a.

The system is of size $48 a_0 \times 88 a_0$. The left $30 a_0 \times 88 a_0$ is the QAHI region (I) and the right $18 a_0 \times L a_0$ is the QAHI+SC region (II), where $L$ is the variable of system length. The rest of the system is filled with vacuum.
Other parameters are chosen as $B = 0.92, a_0=0.59, M=4.46, \alpha=1.75$.
The leads 1 to 3 are assumed to be semi-infinite parabolic metal, attached to the edge of the system through self-energy terms.

\section{Derivation of the edge model\label{app:edge_model}}

We choose the zero-energy solutions of $H_0$ with the form
\begin{align}
    \ket{\gamma_1} &= \frac{1}{\sqrt{2}} (u \ket{e\uparrow} + v \ket{e \downarrow} + u^*\ket{h\uparrow} + v^*\ket{h\downarrow}) \equiv \frac{1}{\sqrt{2}} (\ket{\phi_e} + \ket{\phi_h}) \\
    \ket{\gamma_2} &= \frac{1}{\sqrt{2 i}} (u \ket{e\uparrow} + v \ket{e \downarrow} - u^*\ket{h\uparrow} - v^*\ket{h\downarrow}) \equiv \frac{1}{\sqrt{2 i}} (\ket{\phi_e} - \ket{\phi_h})
\end{align}
where $u$ and $v$ depend on material parameters and the last set of equations define the states $\ket{\phi_{e, h}}$.

The effective edge channel Hamiltonian is constructed in the basis of $\ket{\phi_{e, h}}$. Considering an infinite system, the chiral edge modes with velocity $v_f$ lead to the terms $\bramidket{\phi_e}{H_\textrm{eff}}{\phi_e} = v_f - \mu$, $\bramidket{\phi_h}{H_\textrm{eff}}{\phi_h} = v_f + \mu$.
The pairing term $H_\Delta$ is given by
\begin{equation}
    \bramidket{\phi_e}{H_\Delta}{\phi_h} =
    v^{*2} (d_x+i d_y)-u^{*2}
   (d_x-i d_y)+2 d_z v^* u^*
\end{equation}

The pairing term is determined by the detailed form of $\boldsymbol{d}$ vector. Since the constant paring term $\Delta_0$ cannot appear here due to the particle-hole symmetry, the lowest order term should be linear in $k$. T. Without losing generality, we choose the pairing term as $-v_{\Delta} k$.

\section{Transfer matrix method for transmission\label{app:scattering_matrix}}
In this section, we adopt the transfer matrix method to solve for the transmission $T_\textrm{ee}$ and $T_\textrm{eh}$ given a general effective Hamiltonian $H_\textrm{eff}$, for example Eq.\;(2). This discussion follows Ref.\cite{van2011spin}.

A unitary Hamiltonian needs to be constructed to maintain the conservation of transmission probability. We perform the transformation
\begin{equation}
\tilde{H} = J^{-1/2} H_\textrm{eff} J^{1/2}
\label{eq:H_tilde}
\end{equation}
where $J\equiv \partial H_\textrm{eff} / \partial k$ is the (particle) current operator. The eigenstate of Hamiltonian $\tilde{H}$, denoted by $\tilde{\Psi} \equiv J^{1/2}\Psi$ satisfies $1=\braket{\tilde{\Psi}}{\tilde{\Psi}} = \bramidket{\Psi}{J}{\Psi}$, which means the normalization of $\tilde{\Psi}$ is equivalent to the probability current conversion of $\Psi$. The introduction of $J$ can simplify the expression of probability current conservation.

The wavefunction $\Psi_L$ at position $L$ is related to the initial wavefunction $\Psi_0$ by

\begin{equation}
\Psi_L = Q_L\Psi_0 = W \Lambda_L W^{-1} \Psi_0
\label{eq:Psi_L}
\end{equation}
where $\Lambda_L = \textrm{diag}(e^{i k_1 L}, e^{i k_2 L})$ is the evolution of the eigenstates along the edge. and $W \equiv \{\tilde{\Psi}_1, \tilde{\Psi}_2\}$ is constructed by projecting the incident states onto the eigenstates of $\tilde{H} - \epsilon I$, where $\tilde{\Psi}_i$ is the eigenvector with momentum $k_i$ and energy $\epsilon$. $W$ is the transformation matrix taking account into the interface scattering. The matrix $Q_L$ is the transfer matrix relating the left and right sides of the system.

We consider the initial wavefunction $\Psi_0 = \ket{e}$, as the wavefunction is injected from a QAHI edge state. The transmission is determined entirely by the evolution of the wavefunction through $T=1-2T_\textrm{eh}$ and $T_\textrm{eh} = \left|\bramidket{h}{Q_L}{e}\right|^2$. For the disordered calculations of the single and multiple CEM cases, a total 1000 configurations are used in the average.

\section{Energy and chemical potential dependence of the two models}\label{app:emudependence}
\begin{figure}[htp]
\centering
\includegraphics[width=0.8\columnwidth]{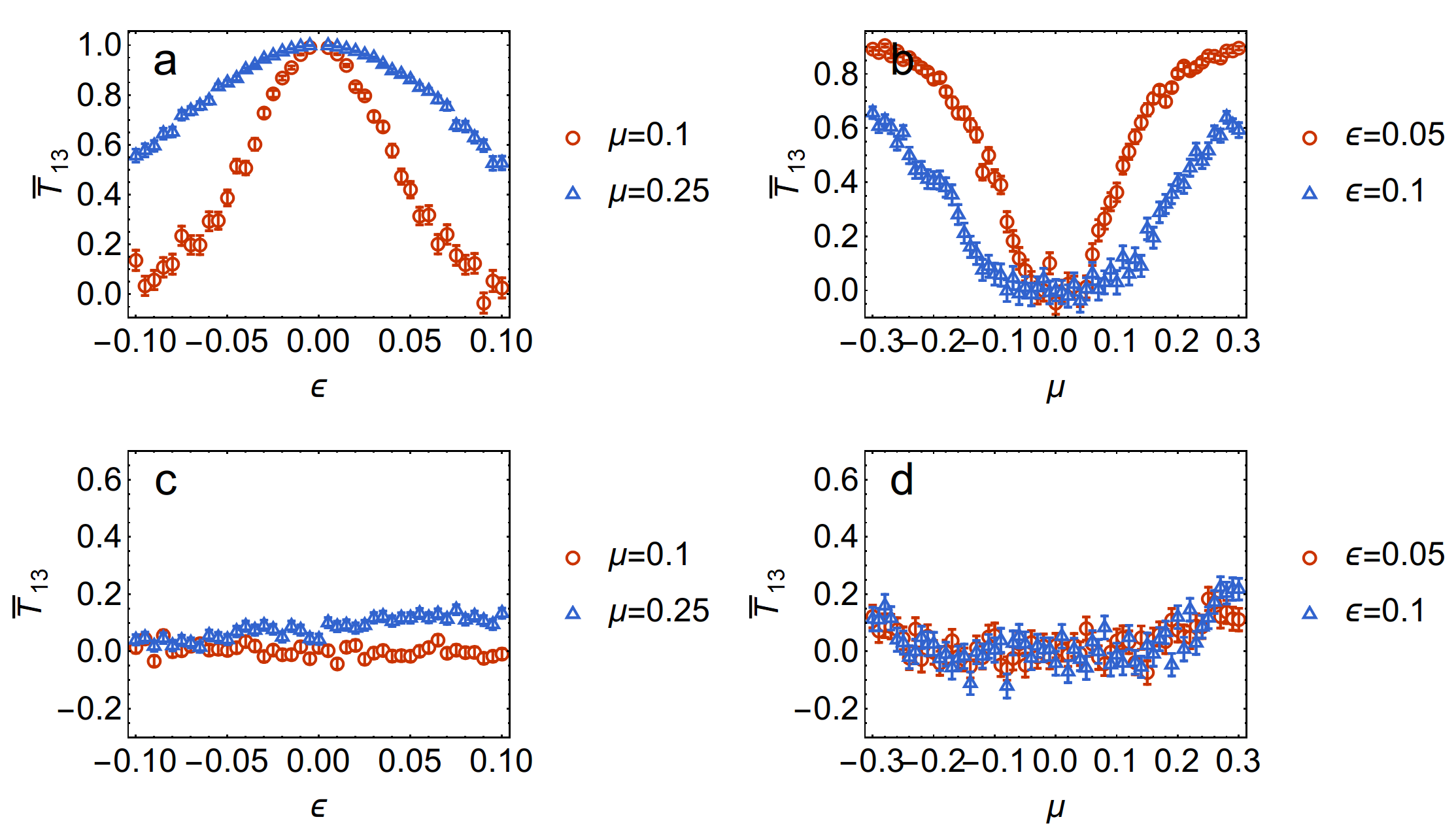}
\caption{(a, b) The energy and chemical potential dependence of the single CEM system. (c, d) Same as (a, b) but for the multiple CEMs system. System length $L=200$. The unitary transmission at zero energy of (a) causes the divergence in resistance in Fig.\;4a.}
\label{fig:emudependence}
\end{figure}

The energy and chemical potential dependence of the single and multiple CEM(s) systems with a fixed length  are shown in \figref{fig:emudependence}. The transmission $\bar{T}_{13} = 1$ at $\epsilon = 0$ for the single CEM case is robust under on-site random potential, as shown in \figref{fig:emudependence}(a). When the absolute value of $\mu$ is decreased, the oscillation for a single disorder configuration is increased, leading to a faster decay to equilibrium when considering the disorder average.

\section{Coexistence of chiral and non-chiral modes \label{app:hybrid_model}}

\begin{figure}[htp]
\centering
\includegraphics[width=0.4\columnwidth]{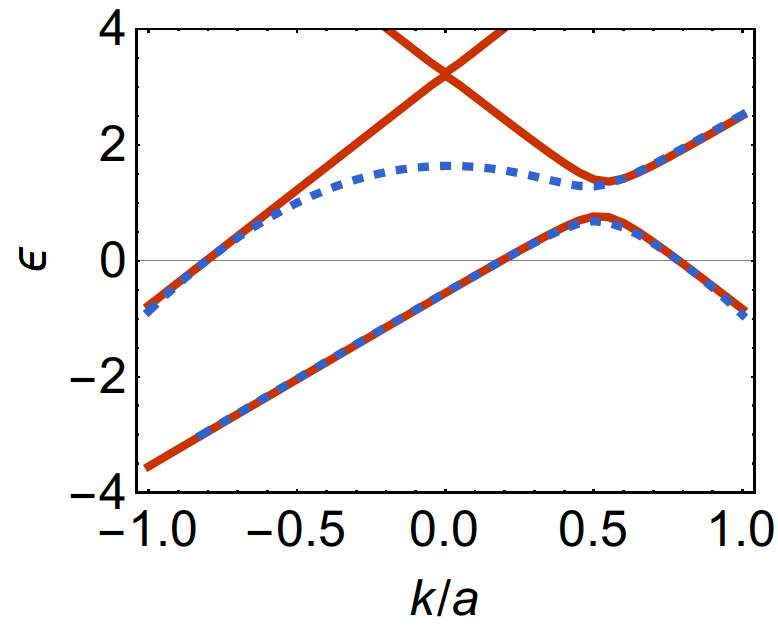}
\caption{(a) Red curves are the band spectra for the Hamiltonian (Eq.\;(6), PH-partner not included) as an example to demonstrate the co-existence of CEM and a pair of helical modes. The non-chiral modes can also approximate the spectrum of a metallic mode with a parabolic dispersion, as shown in blue curves.}
\label{fig:spectrum3band}
\end{figure}

Eq.\;(5) represents a minimal example of a chiral mode coexisting with the non-chiral modes with one left mover and one right mover. The non-chiral modes can originate from the ordinary parabolic band or from the quasi-helical modes at the side surface of topological insulator films. The energy dispersion for this system is shown in Fig. \ref{fig:spectrum3band}.

\begin{figure}[htp]
\centering
\includegraphics[width=0.7\columnwidth]{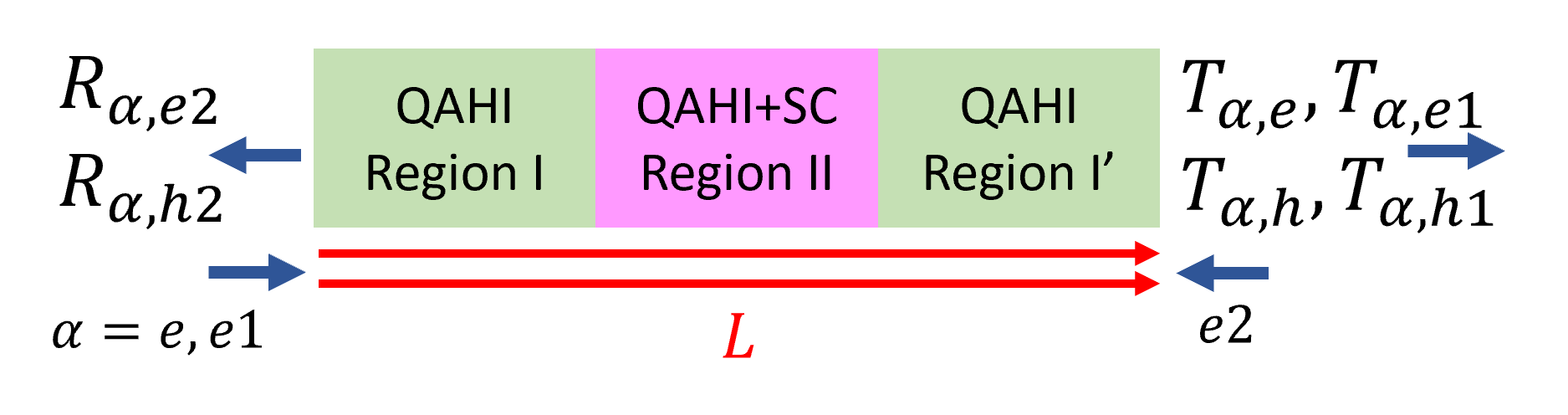}
\includegraphics[width=0.7\columnwidth]{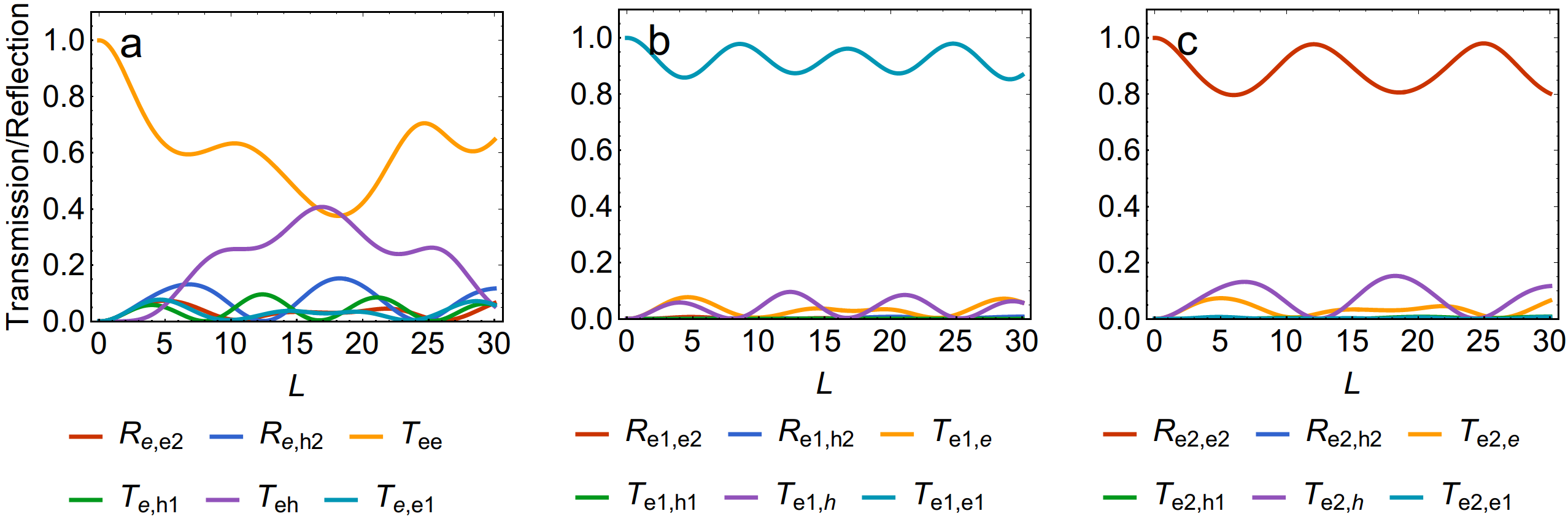}
\caption{A schematic figure of the effective edge model showing the three regions of disordered QAHI region (I and I') and the QAHI+SC region (II). The incident and transmitted/reflected particles are shown in blue arrows. The CEMs are shown in red arrows. (a, b, c) An example of the 18 quantities in \equaref{eq:rawTR6band}, as functions of system length, sorted by the incident mode e, e1 and e2, respectively. The entire length of $L$ is within region II. In (a, b), $e$ and $e1$ are injected at $x=0$, respectively, as indicated by the unitary value of the respective transmissions. For (c), $e2$ is injected at $x=L$.}
\label{fig:rawTR6band}
\end{figure}

The introduction of back-scattered mode causes the left/right boundary and incoming/outgoing modes to not coincide. Thus it is more convenient to switch from the transfer matrix picture into the scattering matrix picture where the matrix relates the incoming and outgoing states. Both the transfer matrix and the scattering matrix are $6 \times 6$ matrices, with 3 electron basis, labeled as $e, e1, e2$, and 3 corresponding hole basis, labeled as $h, h1, h2$. We can transform the $Q$ matrix to the $S$ matrix in the following steps. Firstly, we organize the basis of the Hamiltonian in right-going/left-going blocks and define
\begin{equation}
    \{\Psi_{e\rightarrow L}, \Psi_{h1\rightarrow L}, \Psi_{h\rightarrow L}, \Psi_{e1\rightarrow L}, | \Psi_{e2\leftarrow L}, \Psi_{h2\leftarrow L}\}^T = Q\cdot \{\Psi_{e\rightarrow 0}, \Psi_{h1\rightarrow 0}, \Psi_{h\rightarrow 0}, \Psi_{e1\rightarrow 0}, | \Psi_{e2\leftarrow 0}, \Psi_{h2\leftarrow 0}\}^T
\end{equation}
where the pipe "$|$" separates the right-going and left-going modes. The above equation can be shortened as
\begin{equation}
    \{\Psi_{\rightarrow L}, \Psi_{\leftarrow L}\}^T = Q\cdot \{\Psi_{\rightarrow 0},\Psi_{\leftarrow 0}\}^T
\end{equation}
with $\Psi_{\rightarrow(L, 0)}$ and $\Psi_{\leftarrow(L, 0)}$ being rank 4 and 2 spinors, respectively.

The scattering matrix on the other hand relates incoming and outgoing states, defined as
\begin{equation}
    \{\Psi_{\leftarrow 0}, \Psi_{\rightarrow L}\}^T = S\cdot \{\Psi_{\rightarrow 0},\Psi_{\leftarrow L}\}^T.
\end{equation}
From the above definitions, the relation between the $S$ matrix and the $Q$ matrix is derived as
\begin{equation}
S=\left(
\begin{array}{cc}
 -Q_{\leftarrow\leftarrow}^{-1} Q_{\leftarrow\rightarrow} & Q_{\leftarrow\leftarrow}^{-1} \\
 Q_{\rightarrow\rightarrow} - Q_{\rightarrow\leftarrow}Q_{\leftarrow\leftarrow}^{-1}Q_{\leftarrow\rightarrow} & Q_{\rightarrow\leftarrow}Q_{\leftarrow\leftarrow}^{-1}
\end{array}
\right)
\end{equation}
where the first(second) subscript $\leftarrow$ and $\rightarrow$ correspond to subspaces of $Q$ with dimensions of rows (columns) of 2 and 4, respectively.

The major task is to calculate the the transmission and reflection through the QAH/SC interface, particularly the transmissions and reflections $T_{13}$, $T_{31}$, $R_{11}$ and $R_{33}$ for the leads 1 and 3 in our setup.
Due to the coexistence of chiral and non-chiral modes in our system, we further label the chiral mode as $e$ and $h$ and the non-chiral modes as $e1,e2$ and $h1,h2$ where 1 and 2 denotes two opposite propagation directions, and define the transmissions and reflections between different modes as $R_{\alpha\beta}$ and $T_{\alpha\beta}$ where $\alpha,\beta=e,e1,e2,h,h1,h2$. The transmissions and reflections $T_{ij}$ and $R_{ii}$ ($i,j=1,3$) between different leads can be related to $T_{\alpha\beta}$ and $R_{\alpha\beta}$ by
\begin{equation}
\begin{split}
    T_{13} &= \sum_{\alpha = e, e1} T_{\alpha,e} + T_{\alpha, e1} - T_{\alpha,h} - T_{\alpha,h1} \\
    T_{31} &= T_{e2,e2} - T_{e2,h2}\\
    R_{11} &= T_{e2,e} + T_{e2, e1} - T_{e2,h} - T_{e2,h1}\\
    R_{33} &= \sum_{\alpha = e, e1} R_{\alpha,e2} - R_{\alpha, h2}
\end{split}
\label{eq:6band_transrefle}
\end{equation}
Furthermore, the total probability current conservation requires the conditions
\begin{equation}
\begin{split}
    \sum_{i = e, e1} R_{i,e2}+R_{i,h2}+T_{i,e}+T_{i,h1}+T_{i,h}+T_{i,e1} &= 2\\
    R_{e2,e2}+R_{e2,h2}+T_{e2,e}+T_{e2,h1}+T_{e2,h}+T_{e2,e1} &= 1
\end{split}
\label{eq:rawTR6band}
\end{equation}
\figref{fig:rawTR6band} demonstrates an example of the oscillation behaviors of the 18 quantities on the LHS of \eqref{eq:rawTR6band} as functions of the system length $L$. The unity values at $L=0$ shows the type of incident particle of $e$, $e1$ and $e2$, respectively. The transmissions and reflections of incident holes are associated to the above case by PHS. In the calculation of the total transmission/reflection from the left side, we assume the ensemble of incident electrons are equally distributed between $e$ and $e1$ channels.

\begin{figure}[htp]
\centering
\includegraphics[width=0.75\textwidth]{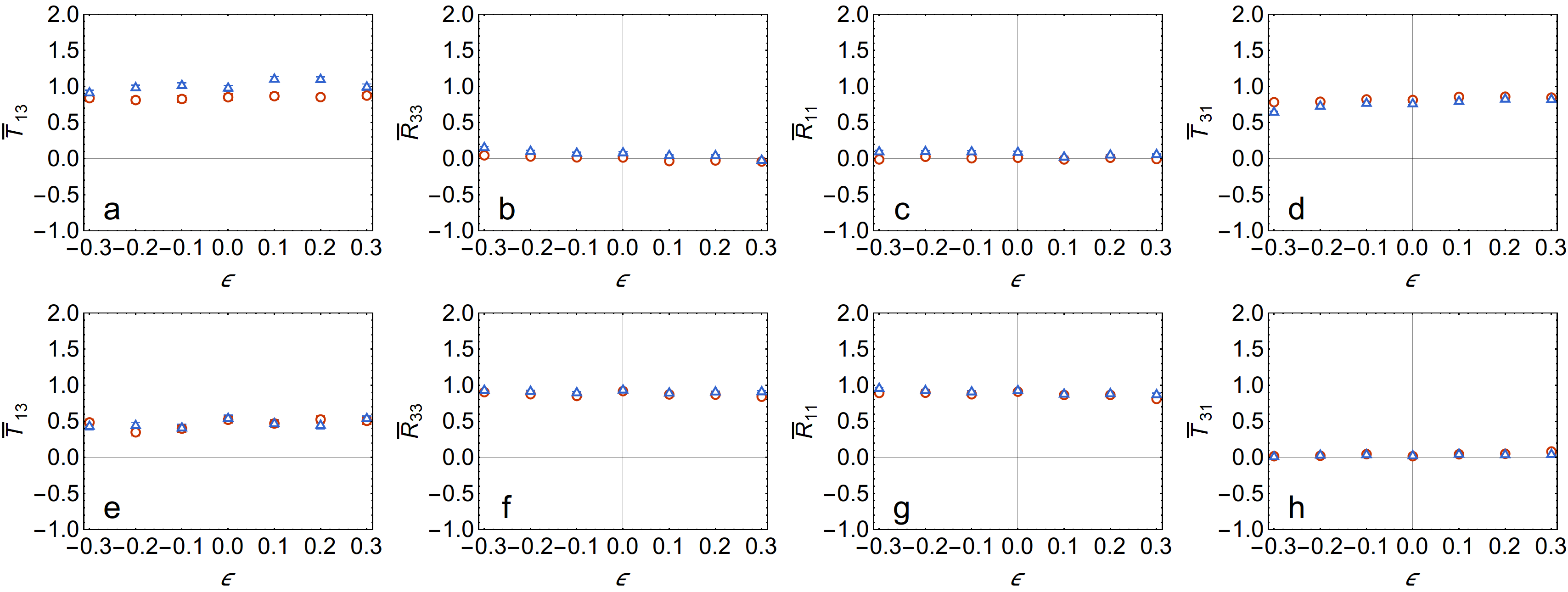}
\caption{The (weak) energy dependence of the forward transmission $\bar{T}_{13}$, reflection $\bar{R}_{33}$ and backward transmission $\bar{T}_{31}$ and reflection $\bar{R}_{11}$ for the coexistence model. (a-d) shows the clean QAH case with system length $L = 200$, (e-h) shows the disordered QAH case with system length $L = 500$. (The values of system length are chosen to be comparable with the decay length of the corresponding systems). Red circles are for parameter set $\mu_1 = 0, \mu_2 = -4.4$, blue triangles are for $\mu_1 = 0.2, \mu_2 = -3.6$. Contrary to the single CEM case, no significant features are present around $\epsilon \sim 0$.}
\label{fig:tModel}
\end{figure}

\figref{fig:tModel}a-d and e-h shows the energy $\epsilon$ dependence of the transmissions and reflections in \equaref{eq:6band_transrefle} for a given system length, for the clean QAH and disordered QAH systems, respectively. The SC region is always disordered for both cases. The weak dependence on $\epsilon$ is the main difference compared with the single CEM case \figref{fig:emudependence}a,b, but this feature is similar to the multiple CEMs case in \figref{fig:emudependence}c,d.

The parameters used for the disordered system simulation are
$v_f = 3, v_{f1} = -v_{f2} = 4.5, k_1 = 1, \xi = 0.3$.
In addition to the disorders on chemical potential $\mu$, an uncorrelated random variation drawn from $[\xi_\text{imp}/2, \xi_\text{imp}/2]$ with $\xi_\text{imp} = 0.3$ is also applied on the electron-coupling term $\xi$ to obtain a shorter decay length for convenience. Qualitatively, we note that the decay and saturation of $\bar{T}_{13}$ and $\bar{R}_{33}$ still holds without the $\xi$ disorder. For both models, a total of 100 configurations are used for deriving the average and standard error. The disordered QAH system of case (iv) is modelled as a one-dimensional system with spatial-dependent $\Delta = 0$ for $x < L_\text{QAH}$ and $x > L_\text{QAH} + L$, where the lengths of the disordered QAH section, $L_\text{QAH}$ is chosen to be 3500, much longer than the disorder localization length, leading to a full Anderson localization for the helical metallic mode, as shown by the schematic plot of the configuration in Fig. \ref{fig:rawTR6band}.

\begin{figure}[htp]
\centering
\includegraphics[width=0.7\columnwidth]{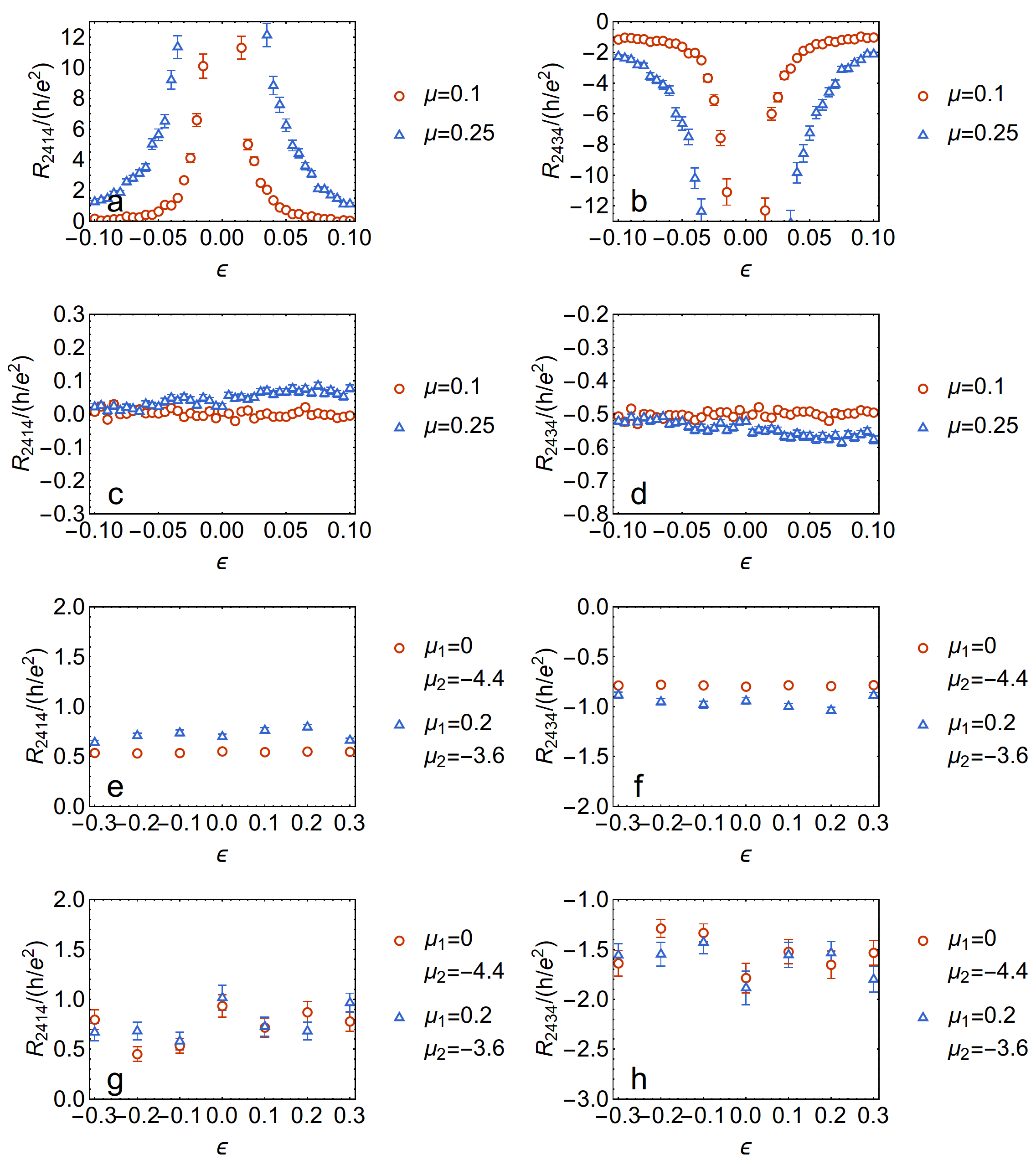}
\caption{Resistances $R_{24,14}$ and $R_{24,34}$ calculated from the transmission (reflection) coefficients, for the four different cases. (a,b) the single CEM case, (c,d) the multiple CEMs case, (e,f) the clean QAH case with coexistence of CEM and metallic mode. (g,h) the disorder QAH case with coexistence of CEM and metallic mode. Subfigure (a, c) are identical to Fig.\;4a,b.}
\label{fig:resistance8}
\end{figure}

In addition to the numerical results shown in the main text, additional calculations are performed to justify some conclusions in the main text. The energy dependence of both $R_{24,14}$ and $R_{24,34}$ of the four cases ((i) single CEM, (ii) multiple CEMs, (iii) clean QAH, (iv) disorder QAH) are shown in \figref{fig:resistance8}, each case with two sets of chemical potential parameters. The relation $R_{24,14} + R_{24,34} = -h/e^2$ for cases (i) and (iv), and $-h/2e^2$ for case (ii) can be verified directly. The quantity $R_{24,14} + R_{24,34}$ for case (iii) remain non-quantized, showing the additional contribution from the non-chiral metallic modes.

\section{Derivation of resistance from the transmission coefficients\label{app:LB}}
In this section, we will derive the resistance of the experimental setup in the Fig.\;1 in main text from the transmissions and reflections for different cases. The transmission coefficients $T$ of the system in Fig.\;1 is written as a $3 \times 3$ matrix, with basis corresponding to lead $1$ to $3$. The current through lead $4$ is implied through total current conservation. The voltage distribution on leads is determined by
\begin{equation}
    \vec{V} = T^{-1}\cdot\vec{I}.
\end{equation}
We consider to drive a current through the lead 2 to the SC (lead 4) and measure the voltage drops at the leads 1 and 3, and thus choose $\vec{I} = \{0, I_0, 0\}$. Consequently, the resistances $R_{24,14} \equiv V_{14}/I_{24}$ and $R_{24,34}\equiv -V_{34}/I_{24}$ can be extracted from the transmission matrix $T$ as follows (we drop the unit $h/e^2$ for the resistance below).
\begin{enumerate}
    \item Single CEM case:
    \begin{align}
        T&=\left(
        \begin{array}{ccc}
         -1 & 0 & T_{13} \\
         1 & -1 & 0 \\
         0 & 1 & -1 \\
        \end{array}
        \right)\\
        R_{24,34} &= \frac{1}{1 - T_{13}}\\
        R_{24,14} &= \frac{T_{13}}{T_{13}-1}
    \end{align}
    \item Multiple CEM case ($N = 2$):
    \begin{align}
        T&=\left(
        \begin{array}{ccc}
         -2 & 0 & 2 T_{13} \\
         2 & -2 & 0 \\
         0 & 2 & -2 \\
        \end{array}
        \right)\\
        R_{24,34} &= \frac{1}{2(1- T_{13})}\\
        R_{24,14} &= \frac{-T_{13}}{2 (1 -T_{13})}
    \end{align}
    \item Coexistence of CEM and non-chiral modes with the clean QAH insulator:
    \begin{align}
        T&=\left(
        \begin{array}{ccc}
         -(3-R_{11}) & 1 & T_{13} \\
         2 & -3 & 1 \\
         T_{31} & 2 & -(3-R_{33}) \\
        \end{array}
        \right)\\
        R_{24,34} &= -\frac{-2 R_{11}+T_{31}+6}{R_{11} (7-3 R_{33})+7 R_{33}+3 T_{13} T_{31}+4 T_{13}+T_{31}-15}\\
        R_{24,14} &= \frac{-R_{33}+2 T_{13}+3}{R_{11} (7-3 R_{33})+7 R_{33}+3 T_{13} T_{31}+4 T_{13}+T_{31}-15}
    \end{align}
    \item Coexistence of CEM and non-chiral modes with the disordered QAH insulator:
    \begin{align}
        T&=\left(
        \begin{array}{ccc}
         -(3-2) & 0 & T_{13} \\
         1 & -(3-2) & 0 \\
         0 & 1 & -(3-1-R_{33}) \\
        \end{array}
        \right)\\
        R_{24,34} &= -\frac{1}{R_{33}+T_{13}-2}\\
        R_{24,14} &= \frac{T_{13}}{R_{33}+T_{13}-2}
    \end{align}
\end{enumerate}
\bibliography{DisorderChiralMajorana}